\documentclass[aps,twocolumn,floats,showpacs]{revtex4}
\usepackage{epsfig}
\usepackage{graphicx}
\usepackage{amsmath}

\newcommand{\beq}{\begin{equation}}
\newcommand{\eeq}{\end{equation}}
\newcommand{\bea}{\begin{eqnarray}}
\newcommand{\eea}{\end{eqnarray}}

\begin{document}

\title{Sufficient conditions for thermal rectification in hybrid quantum structures}
\author{Lian-Ao Wu$^{1,2}$ and Dvira Segal$^{1}$}
\affiliation{$^1$Chemical Physics Group, Department of Chemistry and Center for Quantum
Information and Quantum Control, University of Toronto, 80 St. George
street, Toronto, Ontario, M5S 3H6, Canada}
\affiliation{
$^2$Department of Theoretical Physics and History of Science,
The Basque Country University (EHU/UPV), PO Box 644, 48080 Bilbao, Spain}
\date{\today}

\begin{abstract}
We analytically identify  sufficient conditions for manifesting
thermal rectification in  two-terminal hybrid structures within the
quantum master equation formalism. We recognize two classes of
rectifiers. In type A rectifiers the contacts are dissimilar. In
type B rectifiers the contacts are equivalent, but the system and
baths have different particle statistics, and the system is
(parametrically) asymmetrically coupled to the  baths. Our study
applies to various hybrid junctions including metals, dielectrics,
and spins.
\end{abstract}

\maketitle


Understanding heat transfer in  hybrid structures is of fundamental
and practical importance for controlling transport at the nanoscale,
and for realizing functional devices \cite{Majumdar-rev}. Among the
systems that fall into this category are metal-molecule-metal
junctions, the basic component of molecular electronic devices
\cite{MolEl}, and dielectric-molecule-dielectric systems, where
vibrational energy flow activates reactivity and controls dynamics
\cite{Dlott}. Phononic junctions are also captivating and essential
for understanding the validity of the Fourier's law of thermal
conduction at the nanoscale \cite{Fourier,ZettlF}.
Single-mode {\it radiative} heat conduction between ohmic metals was
recently detected, showing that photonic thermal conductance is
quantized \cite{Pekola}. Other hybrid systems with interesting
thermal properties are electronic spin-nuclear spin interfaces
\cite{Lukin}, metal-molecule contacts with exciton to phonon energy transfer \cite{Backus},  
and metal-superconductor junctions \cite{Rev}.

\begin{figure}[htbp]
\hspace{2mm}
{\hbox{\epsfxsize=80mm \epsffile{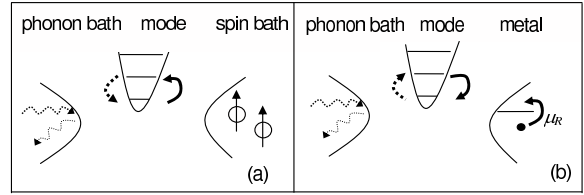}}}
\caption{Examples of two hybrid systems treated in this work.
(a) Single mode heat transfer between a solid and a spin bath.
(b) Phonon to exciton energy exchange. 
The central unit can represent either a vibrational  or a radiation  mode.}
\label{Fig0}
\end{figure}

Thermal rectification, namely an asymmetry of the heat current for
forward and reversed temperature gradients, has recently attracted
considerable theoretical
\cite{Terraneo,Casati,Rectif,Zhang,Prosen,RectifMass,Zeng} and
experimental \cite{RectifE,QDrectif} attention. Most theoretical
studies, confined to a specific realization, have analyzed this
phenomenon in {\it phononic systems} using classical molecular
dynamics simulations. In this letter we attempt a first step towards an
analytical understanding of this effect.
We  establish
sufficient conditions for manifesting thermal rectification in a
prototype-hybrid quantum model, including a central quantized unit
(subsystem) and two bulk objects (reservoirs). We identify two
classes of thermal rectifiers: (i) Type A rectifiers where the
terminals are dissimilar i.e. of different mean energy (or heat
capacity). (ii) Type B rectifiers, where the contacts are
equivalent, but the reservoirs and subsystem have different
statistics, combined with unequal coupling strengths at the two
ends.
We manifest that these rectifiers could be realized in several subsystems (harmonic and anharmonic) and reservoirs
(spin, metal, dielectrics), see Fig. \ref{Fig0}.

Consider a 1-dimensional  hybrid structure where a central unit $H_S$
interacts with two reservoirs $H_{\nu}^0$  ($\nu=L,R$) of
temperatures $T_{\nu}=\beta_{\nu}^{-1}$ via the coupling terms $V_{\nu}$,
\bea
H=H_L^0+H_R^0+H_S+V_L+V_R.
\label{eq:H}
\eea
The heat current from the left bath into the subsystem is given by
$J_{L}=\frac{i}{2}{\rm Tr}\left([H_{L}^{0}-H_{S},V_{L}]\rho \right)$; ($\hbar\equiv 1$)
 \cite{Current},
where $\rho$ is the total density matrix, and we trace over the system and
reservoirs degrees of freedom. In steady-state the expectation
value of the interaction is zero, ${\rm Tr}\left(\frac{\partial
V_{L}}{\partial t}\rho \right)=0$, and we obtain
\bea
J=i{\rm Tr} [\widehat J \rho]; \,\,\,\,\,\,
\widehat{J}=\frac{i}{2}[V_{L},H_{S}]+\frac{i}{2}[H_{S},V_{R}].
\label{eq:JL}
\eea
This expression was derived  based on the equality $J_L=-J_R$ in steady-state.
We discard the subscript $L$ in (\ref{eq:JL}), thus the heat current
is defined positive when flowing left to right.
The system Hamiltonian assumes a diagonal form, and we also consider separable couplings
\bea
H_{S}&=&\sum_n
E_{n}|n\rangle \langle n|;
\nonumber\\
V_{\nu}&=&SB_{\nu}; \,\,\,\,S=\sum_{n,m} S_{m,n}|m\rangle \langle n|.
\label{eq:HSV}
\eea
Here $S$ is a subsystem operator and $B_{\nu}$ is an operator in
terms of the $\nu$ bath degrees of freedom. For simplicity we set
$S_{m,n}=S_{n,m}$. 
In what follows we consider cases where
$B_{L}$ and $B_R$ have equal structure but different prefactors.
We refer to this case as "parametric asymmetry", or "unequal coupling strength",
rather than "functional asymmetry", resulting from dissimilar $B$'s.
Note that if the commutator $[H_S,S]=0$,  the heat current trivially vanishes.

We begin and discuss the type A  rectifier,  constructed by adopting reservoirs with distinct properties, as we explain below.
First we derive a general expression for the heat flux in hybrid structures.
Formally, if we write the heat current
as $J(T_a,\Delta)=\sum_{k}\alpha_k(T_a)\Delta^k$, rectification takes place if even terms survive,
$\alpha_{k=2n}\neq 0$; $n=1,2...$, $T_a=T_L+T_R$ and $\Delta=T_L-T_R$.

The initial density matrix $\rho$ is  assumed to be a tensor product of
system $\rho_S$ and bath $\rho_B=\rho_L(T_L) \otimes \rho_R(T_R)$ factors, 
where $\rho_{\nu}(T)=e^{-\frac{H_{\nu}^0}{T}}/{\rm Tr_{\nu}}\big[e^{-\frac{H_{\nu}^0}{T}}\big]$ ($\nu=L,R$).
For convenience we delete the direct reference to time.
In terms of $T_a$ and $\Delta$ we can write
\bea
\rho(T_a,\Delta) 
=\frac{1}{Z} e^{-\frac{ 2T_a (H_L^0 + H_R^0) -2\Delta (H_L^0 - H_R^0)    }{T_a^{2}-\Delta ^{2}} }
\otimes \rho_S,
\label{eq:rho}
\eea
where  $Z={\rm Tr}[\rho_S \rho_B]$ is the partition function with  the trace  performed
over bath and system degrees of freedom.
Furthermore, without loss of generality, 
the system density matrix 
is assumed to depend initially only on the average temperature, $\rho_S\equiv\rho_S^{(0)}(T_a/2)$.

The expectation value of the energy current is given by evaluating
$J(T_a,\Delta) =\text{Tr}[\widehat{J}\rho (T_a,\Delta )]$,  where $\widehat J$ is an Heisenberg
representation operator $\widehat J=e^{iHt}\widehat J(0) e^{-iHt}$.
 Expanded in powers of $\Delta$ we get
\bea
J(T_a,\Delta )=\frac{2\Delta}{T_a^{2}}\alpha_0(T_a) + \frac{(2\Delta) ^2}{T_a^{4}} \alpha_1(T_a)
+O\left( T_a,\Delta ^{3}\right),
\label{eq:exp}
\eea
with the coefficients
\bea
\alpha_0(T_a) &=&\text{Tr}\left[ \widehat{J}(H_{L}^{0}-H_{R}^{0})\rho_{T}\right],
\label{eq:alpha0}
\\
\alpha_1(T_a) &=&\frac{1}{2}\text{Tr}\left[ \widehat{J}(H_{L}^{0}-H_{R}^{0})^{2}\rho_T
\right]
\nonumber \\
&-&\alpha_0(T_a)
\text{Tr}\left[ (H_{L}^{0}-H_{R}^{0})\rho _{T}\right] .
\label{eq:alpha1}
\eea
Here $\rho_{T}= \frac{1}{Z_{T_a}}\rho_L(\frac{T_a}{2}) \rho_R(\frac{T_a}{2}) \rho_S^{(0)}(\frac{T_a}{2})$ is the density matrix at
the average temperature $\frac{T_L+T_R}{2}$ with the partition function
$Z_{T_a}$. 
Note that the operators are time dependent, given in their Heisenberg representation. 
Eq. (\ref{eq:alpha1}) was derived using the fact that Tr$\left[ \widehat{J}\rho _{T}\right]$ and all its
$T$ derivatives are zero.
Using the definition of the current operator (\ref{eq:JL}) we obtain an explicit expression for $\alpha_0$,
\bea
\alpha_0(T_a)= i\text{Tr} \big\{[S,H_S](B_LH_L^0 +B_R H_R^0)\rho_T \big\}.
\eea
Since $[H_S,S]\neq 0$ [see discussion after Eq. (\ref{eq:HSV})], the linear  term in $J$ is  finite \cite{comment}.

We now examine the onset of thermal rectification, i.e. discuss the sufficient conditions for having $\alpha_1\neq 0$.
While there might be some special values of $T_a$ where the two terms in (\ref{eq:alpha1})
cancel, in general since $B_{\nu}$ and $H_{\nu}^0$ are independent operators, the result is finite.
Thus, in order to manifest rectification it is enough to analyze when one of the
terms in (\ref{eq:alpha1}) is nonzero.
(i) The first expression is finite if $\text{Tr}[F H_L^0 \rho_T]\neq \text{Tr}[FH_R^0 \rho_T]$;
$F= [S,H_S](B_LH_L^0 +B_R H_R^0)$.
The second term is nonzero if  $\text{Tr}[H_L^0\rho_T] \neq \text{Tr}[H_R^0 \rho_T]$. 
Based on (ii) we conclude that  rectification emerges
when the reservoirs have different mean energy,
\bea
\langle H_L^0\rangle\equiv  {\rm Tr}_L[ \rho_L(T) H_L^0 ] \neq {\rm Tr}_R[ \rho_R(T) H_R^0 ] \equiv \langle H_R^0 \rangle.
\eea
As an example, consider a bath of  1-dimensional oscillators,
$H_{L}^0=H_L^{kin}+H_L^{pot}$, where the kinetic energy $H_L^{kin}$
is quadratic in momentum and the potential energy per particle is
$C_nq^n$; $n\geq 2$. In the classical limit using the equipartition
relation we obtain $\langle H_L^0 \rangle =
T_L(\frac{1}{2}+\frac{1}{n})$. Thermal rectification thus emerges if
the reservoirs have a non-identical power $n$.
%
Note that the separation to three segments ($L$, subsystem, $R$) is often artificial,
as the system can be practically made of a single structure  with a varying potential energy, e.g. 
an asymmetrically  mass loaded nanotube \cite{RectifE}.
Our results manifest that such an inhomogeneous structure should rectify heat.
Finally, we comment that our discussion could be generalized to cases where $V_{\nu}=S_{\nu}B_{\nu}$;
$S_L\neq S_R$, see Eq. (\ref{eq:HSV}).

We turn to the type B rectifier,
and show that for equivalent reservoirs rectification emerges
when the subsystem  and reservoirs have different statistics, in conjunction with some parametric asymmetry.
It is easy to show that under (\ref{eq:HSV}) the steady-state current (\ref{eq:JL}) becomes
%
$J=i\sum_{n,m} E_{m,n}S_{m,n}{\rm Tr_{B}}(B_{L}\rho _{m,n})$,
%
where $E_{m,n}=E_m-E_n$, and ${\rm Tr_B}$ denotes the trace over the $L$ and $R$ reservoirs degrees of freedom.
Employing the Liouville equation in the interaction picture,
the elements of the total density matrix satisfy
\bea
\frac{d\rho _{m,n}}{dt}= -\int_0^t d\tau [V(t),[V(\tau),\rho(\tau)]]_{m,n}
\label{eq:Lioville}
\eea
where $V=V_L+V_R$, and $V(t)$ are interaction picture operators.
Following the standard weak coupling scheme \cite{Breuer}, going to the markovian limit, the
heat current reduces to
\bea
J=\frac{1}{2}\sum_{n,m} E_{m,n}\left\vert S_{m,n}\right\vert
^{2}P_{n}\times (k_{n\rightarrow m}^{L}-k_{n\rightarrow m}^{R}),
\label{eq:current}
\eea
where the transition rates are  given by
%
$k_{n\rightarrow m}^{\nu}=\int_{-\infty }^{\infty }d\tau
e^{iE_{n,m}\tau }\left\langle B_{\nu}(\tau )B_{\nu}(0)\right\rangle$,
%
and the population $P_n={\rm Tr_B}(\rho_{n,n})$ satisfies the differential equation
\bea
\dot P_n=\sum_{\nu,m}|S_{m,n}|^2P_m k_{m \rightarrow n}^{\nu} -P_n \sum_{\nu,m} |S_{m,n}|^2
k_{n\rightarrow m}^{\nu}.
\label{eq:master}
\eea
In steady-state $\dot P_n=0$, and we normalize the population to unity
$\sum_n P_n=1$.
Our description to this point is general, as we have not yet specified neither the subsystem
nor the interfaces. 
We consider next two representative models for the system Hamiltonian and its interaction with the baths.
In the first model the subsystem is a harmonic oscillator (HO) of frequency $\omega$,
$H_S=\sum_n n\omega|n\rangle \langle n|$.
This can describe either a local radiation mode \cite{Pekola,radiation} or a
vibrational mode of the trapped molecule \cite{NDR}.
We also take $S=\sum_n\sqrt{n} |n\rangle \langle n-1| + c.c$,
motivated by the bilinear form $V_{\nu}\propto xB_{\nu}$,  $x$ is a subsystem coordinate \cite{NDR}.
This implies that only transitions between nearest states are allowed,
\bea
&&k^{\nu}\equiv k_{n\rightarrow n-1}^{\nu}=\int_{-\infty }^{\infty }d\tau e^{i\omega
\tau }\left\langle B_{\nu}(\tau )B_{\nu}(0)\right\rangle;
\nonumber\\
&&k_{n-1\rightarrow n}^{\nu}=e^{-\beta_{\nu}\omega}k^{\nu}_{n\rightarrow n-1}.
\label{eq:raten}
\eea
Solving (\ref{eq:master}) in steady-state using the rates (\ref{eq:raten}),
the heat current (\ref{eq:current}) can be analytically calculated,
\bea
J^{(HO)}=-\frac{\omega \lbrack n_{B}^{L}(\omega )-n_{B}^{R}(\omega )]}
{n_{B}^{L}(-\omega )/k^{L}+n_{B}^{R}(-\omega )/k^{R}},
\label{eq:HO}
\eea
where $n_{B}^{\nu}(\omega )=\left[e^{\beta_{\nu}\omega}-1\right]^{-1}$
is the Bose-Einstein distribution function at $T_{\nu}=1/\beta_{\nu}$.

Our second subsystem is a two-level system (TLS).
Here $H_S=\frac{\omega}{2}\sigma_z$, and we employ a nondiagonal interaction  $S=\sigma_x$.
These terms can represent an electronic spin rotated by the environment
\cite{Lukin}. It can also describe an anharmonic (truncated) molecular vibration
dominating  heat flow through the junction \cite{Rectif,NDR}.
Re-calculating the long-time population (\ref{eq:master}),
the heat flux reduces to
\bea
J^{(TLS)}=\frac{\omega \lbrack n_{S}^{L}(\omega )-n_{S}^{R}(\omega )]}{%
n_{S}^{L}(-\omega )/k^{L}+n_{S}^{R}(-\omega )/k^{R}}
\label{eq:TLS}
\eea
with the rates (\ref{eq:raten}) and the spin occupation factor
$n_{S}^{\nu}(\omega)=\left[e^{\beta_{\nu} \omega}+1\right]^{-1}$.
Expressions (\ref{eq:HO}) and (\ref{eq:TLS}) show that
in the weak coupling limit
the effect of the environment enters only through the relaxation rates $k^{\nu}$, evaluated
at the subsystem energy spacing $\omega$.

We now analyze the general structure of the last two expressions,
and discuss the onset of thermal rectification.
It is clear that if $k^L(T)=k^R(T)$, i.e.
the reservoirs and system-bath interactions $B_{\nu}$ are equivalent, thermal rectification is absent
as $J(\Delta)=-J(-\Delta)$.
On the other hand, if $k^L(T)=f(T)k^R(T)$, resulting e.g. from the use of dissimilar  reservoirs,
the system generally rectifies heat besides
some special points in the parameter space, depending on the details of the model.
This case reduces to the type A rectifier discussed above.

However, a more careful analysis of Eqs. (\ref{eq:HO}) and (\ref{eq:TLS})
reveals that rectification prevails if $k^L(T)=ck^R(T)$,  $c\neq 1$ is a constant,
{\it  given that the relaxation rates' temperature dependence differs from the central unit
particle statistics.} 
For example, for a spin (TLS) subsystem  we require that $n_S^{\nu}(-\omega)/k^{\nu}=g(T_{\nu})$, a non-constant function
of temperature. A TLS asymmetrically coupled to two
harmonic baths thus rectifies heat \cite{Rectif}. 
As we show next, the temperature dependence  of the rates $k^{\nu}$ reflects
the reservoirs statistics. We therefore classify
 type B rectifiers as junctions  where the system and bath differ in their statistics, and
the equivalent reservoirs are (parametrically) asymmetrically coupled
to the system.
We specify next the contacts, and exemplify the two classes of 
thermal rectifiers in various hybrid structures.

\textit{Spin bath}. Assuming the environment includes a set of
 distinguishable noninteracting spin-$1/2$ particles
($p=1,2,..,P$),
the bath Hamiltonian is given by summing all separable contributions,
$H_{\nu}^0=\sum_p h^0_{\nu,p}$ and $B_{\nu}=\sum_p b_{\nu,p}$.
The relaxation rate (\ref{eq:raten}) reduces to
\bea
k^{\nu}=n_{S}^{\nu}(-\omega )\Upsilon_{\nu}(\omega),
\label{eq:R1s}
\eea
where 
 $n_{S}^{\nu}(\omega)$=$\left[e^{\beta_{\nu} \omega}+1\right]^{-1}$ and
$\Upsilon_{\nu}(\omega )=2\pi\sum_{p}\left\vert \left\langle 0\right\vert
_{p}b_{\nu,p}\left\vert 1\right\rangle _{p}\right\vert ^{2}\delta (\omega+\epsilon_p(0)
-\epsilon_{p}(1))$,
with the $p$-particle eigenstates $\left\vert i\right\rangle _{p}$ and eigenvalues $\epsilon_{p}(i)$ ($i=0,1$).

\textit{Solid/Radiation field (harmonic bath)}. This bath includes a
set of independent harmonic oscillators, creation operator
$a_{\nu,j}^{\dagger}$. System-bath interaction is further assumed to
be bilinear, $H_{\nu}^0=\sum_{j}\omega_j a_{\nu,j}^{\dagger}
a_{\nu,j}$; $B_{\nu}=\sum_{j}
\lambda_{\nu,j}(a_{\nu,j}+a_{\nu,j}^{\dagger })$, where
$\lambda_{\nu,j}$ are the system-bath coupling elements. This leads
to the relaxation rate (\ref{eq:raten})
\bea k^{\nu}=-\Gamma_B^{\nu}(\omega)n_{B}^{\nu}(-\omega),
\label{eq:R2} \eea
where $\Gamma_B^{\nu}(\omega)=2\pi \sum_j \left\vert
\lambda_{\nu,j}\right\vert ^{2}\delta (\omega _{j}-\omega )$ is an
effective system-bath coupling energy and $n_B^{\nu}$ is the Bose-Einstein distribution function
at temperature $T_{\nu}$.

\textit{Metal}.
As a final example the contact is made metallic,
including a set of noninteracting spinless electrons, creation operator $c_{\nu,i}^{\dagger}$,
The bath operator coupled to the system allows scattering between electronic states within the same lead,
$H_{\nu}^0=\sum_i \epsilon _{i}c_{\nu,i}^{\dagger }c_{\nu,i}$;
$B_{\nu}=\sum_{i,j} v_{\nu,i,j}c_{\nu,i}^{\dagger }c_{\nu,j}$.
The transition rate (\ref{eq:raten}) can be written as
\bea
k^{\nu} &=&-2\pi n_{B}^{\nu}(-\omega ) \sum_{i,j} \left\vert v_{\nu,i,j}\right\vert ^{2}\delta (\epsilon
_{i}-\epsilon _{j}+\omega )
\nonumber\\
&&\times \lbrack n_{F}^{\nu}(\epsilon_{i})-n_{F}^{\nu}(\epsilon _{i}+\omega )],
\eea
with the Fermi-Dirac distribution function $n_F^{\nu}(\epsilon)=[e^{\beta_{\nu}(\epsilon-\mu_{\nu})}+1]^{-1}$ at
the chemical potential
$\mu_{\nu}$.
One could also write
\bea
k^{\nu}=-n_{B}^{\nu}(-\omega )\Lambda^{\nu}(T_{\nu},\omega),
\label{eq:R4}
\eea
where
$\Lambda^{\nu}(T_{\nu,}\omega)=
2\pi \int d\epsilon \left[ n_{F}^{\nu}(\epsilon )-n_{F}^{\nu}(\epsilon +\omega )\right]F_{\nu}(\epsilon)$.
The function $F_{\nu}(\epsilon)=\sum \left\vert
v_{\nu,i,j}\right\vert ^{2}\delta (\epsilon -\epsilon _{j}+\omega
)\delta (\epsilon _{i}-\epsilon )$ depends on the system-bath
coupling elements and the specific band structure. Assuming that the
density of states slowly varies in the  energy window $\omega$, this
function could be expanded around the chemical potential
\cite{radiation}. If the Fermi energy is much bigger than the
conduction band edge  we  obtain 
%
$\Lambda^{\nu}(T_{\nu,}\omega)\approx \Gamma_F^{\nu} \left(1+\delta_{\nu} \frac{T_{\nu}}{\mu_{\nu} }\right)$,
%
where $\Gamma_{F}^{\nu}= 2\pi\omega F_{\nu}(\mu_{\nu})$,
and $\delta$ is a constant of order one, measuring the deviation from a flat band structure near the chemical potential \cite{radiation}.


\begin{figure}[htbp]
\hspace{2mm}
{\hbox{\epsfxsize=70mm \epsffile{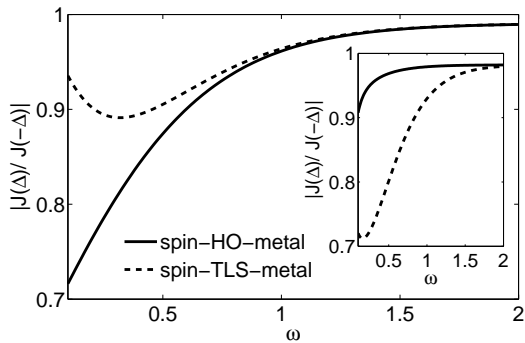}}}
\caption{Spin-HO-Metal rectifier (full) and a Spin-TLS-Metal rectifier (dashed).
Rectification ratio is presented as a function of the subsystem spacing.
$T_a=0.5$, $\Delta=0.1$, $\delta_{R}=0.2$, $\mu_{R}=1$,
Main plot: The subsystem is equally coupled to the two ends, $\Upsilon_L=\Gamma_F^R=1$.
Inset: The effect can be tuned by manipulating system-bath couplings, $\Upsilon_L=1$, $\Gamma_F^R=0.05$.
}
\label{Fig1}
\end{figure}

Consider for example a type A solid-HO-metal rectifier, representing
an electronic to vibrational energy conversion device, see Fig.
\ref{Fig0}(b). This system may be realized by attaching an
insulating molecule to a metal (STM tip), while the underneath
surface is insulating. Setting the coupling strength at both
contacts to be the same, $\Gamma_B^{L}(\omega )=\Gamma_F^R
(\omega)$, we can calculate the rectification ratio ${\mathcal
R}\equiv \left\vert \frac{J(\Delta )}{J(-\Delta )}\right\vert$ using
(\ref{eq:HO}), (\ref{eq:R2}) and (\ref{eq:R4}) 
\bea
{\mathcal R}=\frac{(2+\delta_R \frac{T_L}{\mu_R})(1 +\delta_R \frac{T_R}{\mu_R})}
{(1 +\delta_{R}\frac{ T_L}{\mu_R})(2 +\delta_R \frac{T_R}{\mu_R})}
\sim 1-\Delta\frac{\delta_R}{2\mu_R}. \eea
Therefore, if the metal density of state varies with energy
($\delta\neq0$), thermal rectification is presented
\cite{radiation}. Interestingly, we can show that in a
solid-TLS-metal junction
 $\mathcal R$  could be modulated to be greater or smaller than one
 by varying the gap $\omega$.
Thus, phonon-to-exciton heat conversion can be made effective,
while the exciton-to-phonon route becomes ineffective, and vice versa.
As a second- type B- example consider  a solid-HO-solid structure.
In the classical limit [see (\ref{eq:TLS}) and (\ref{eq:R2})] we get
\bea
J(T_a,\Delta)&=&\frac{\omega\Gamma_B^L\Gamma_B^R\Delta}{2\Gamma_B^L T_L+ 2\Gamma_B^RT_R}
\nonumber\\
&\propto& \frac{\Delta}{T_a}\left(1-x\frac{\Delta}{T_a}+x^2\frac{\Delta^2}{T_a^2}+...\right),
\eea
with $x=(\Gamma_B^L-\Gamma_B^R)/(\Gamma_B^L+\Gamma_B^R)$.
This demonstrates that even $\Delta$ terms, i.e. thermal
rectification, are directly linked to the parametric asymmetry.
Fig. \ref{Fig1} further displays the tunability  of a
spin-subsystem-metal junction. In the classical limit
($\omega<T_{\nu}$) rectification can be substantial, while in the
quantum regime the effect is suppressed. Modifying the system-metal
coupling strength  largely controls the rectification ratio (inset).



To summarize, while  previous studies  
were focused on a specific realization, typically limited to the
classical regime, based on numerical simulations, we have
analytically deduced sufficient condition for the onset of thermal
rectification in generic hybrid structures at the level of the
quantum master equation: (i) The reservoirs should be made
dissimilar, e.g. rectification emerges in an anharmonic junction
where the $L$ and $R$ segments have different potentials. (ii) The
contacts could be of the same type, but their statistics should differ from that of the
system, combined with some parametric asymmetry, e.g. a 
boson-spin-boson junction rectifies heat when $\Gamma_B^L\neq
\Gamma_B^R$.
Our study applies to various interfaces: metals, insulators and noninteracting spins.
The central unit could represent  a radiation mode, a vibrational mode, or an electronic excitation.

Anharmonic interactions were in particular pointed out responsible for thermal rectification.
While it is obvious that perfectly harmonic systems cannot rectify heat \cite{Fourier},
not all anharmonic-asymmetric systems do bring in the effect.
Consider for example  a three-segment nonlinear oscillators chain where
all units have identical potentials, but the central part is asymmetrically  connected to the two terminals.
According to our analysis this system will not rectify heat.
%

%
Our study manifests control over energy transfer at the nanoscale,
important for cooling electronic and mechanical devices and for controlling molecular reactivity.
We also demonstrated that thermal rectification is an abundant effect that could be observed in
a variety of systems, phononic \cite{RectifE},  electronic \cite{QDrectif}, and
photonic \cite{Pekola,radiation}.

\vspace{0.2in} \textbf{Acknowledgement} This work was supported by
the University of Toronto Start-up Funds.



\begin{thebibliography}{99}


\bibitem{Majumdar-rev}
V. P. Carey {\it et al.},
Nanoscale and Microscale Thermophysical Engineering {\bf 12}, 1 (2008).



\bibitem{MolEl}
A. Nitzan and M. A. Ratner,
Science, {\bf 300}, 1384 (2003).

%
\bibitem{Dlott}
Z. Wang {\it et al.}
Science {\bf 317}, 787 (2007).

\bibitem{Fourier}
F. Bonetto, J. Lebowitz, and L. Rey-Bellet, {\it Mathematical Physics
2000} (World Scientific, Singapore, 2000), pp. 128–150.

\bibitem{ZettlF}
C. W. Chang {\it et al.},
Phys. Rev. Lett. {\bf 101}, 075903 (2008).

\bibitem{Pekola}
M. Meschke, W. Guichard, and J. P. Pekola,
Science {\bf 444}, 187 (2006).

\bibitem{Lukin}
J. M. Taylor, C. M. Marcus, and  M. D. Lukin,
Phys. Rev. Lett. {\bf 90}, 206803 (2003).


%
\bibitem{Backus}
E. H. G. Backus {\it et al.},
Science {\bf 310}, 1790 (2005).

\bibitem{Rev}
F. Giazotto, {\it et al.}
Rev. Mod. Phys. {\bf 78}, 217 (2006).










\bibitem{Terraneo}
M. Terraneo, M. Peyrard, and  G. Casati,
Phys. Rev. Lett. {\bf 88}, 094302 (2002).

\bibitem{Casati}
B.  Li, L. Wang, and. G. Casati,
Phys. Rev. Lett. {\bf 93}, 184301 (2004).

\bibitem{Rectif}
D. Segal and A. Nitzan, Phys. Rev. Lett.  {\bf 94}, 034301 (2005);
 J. Chem. Phys. {\bf 122}, 194704  (2005).


\bibitem{Zhang}
B. Hu, L. Yang, and Y. Zhang,
Phys. Rev. Lett. {\bf 97}, 124302 (2006).

\bibitem{Prosen}
G. Casati, C. Mejia-Monasterio, and T. Prosen,
Phys. Rev. Lett. {\bf 98}, 104302 (2007).



\bibitem{RectifMass}
N. Yang, N. Li, L. Wang, and B. Li,
Phys. Rev. B {\bf 76}, 020301 (2007).

\bibitem{Zeng}
N. Zeng and J.-S. Wang, Phys. Rev. B {\bf 78}, 024305 (2008).



\bibitem{RectifE}
C. W. Chang {\it et al.},
Science {\bf 314},
1121 (2006).

\bibitem{QDrectif}
R. Scheibner {\it et al.},
New. J. Phys. {\bf 10 },  083016 (2008).


\bibitem{Current}
L.-A. Wu and D. Segal,     arXiv:0804.3371v1.




\bibitem{comment}
All even $\Delta$ terms of $J$ have a contribution from  the preceding odd term times 
$\text{Tr}\left[ (H_{L}^{0}-H_{R}^{0})\rho_{T}\right]$. The following discussion
thus holds for higher orders as well.


\bibitem{Breuer}
H.-P. Breuer and F. Petruccione,
{\it The Theory of Open Quantum Systems},
Oxford University Press, New York, New York, (2002),


\bibitem{radiation}
D. Segal, Phys. Rev. Lett. {\bf 100}, 105901  (2008).

\bibitem{NDR}
D. Segal, Phys. Rev. B {\bf 73}, 205415 (2006).











\end{thebibliography}
\end{document}